\documentclass[11pt, a4paper]{article}
\usepackage{amsmath}
\usepackage{amssymb}
\usepackage[dvips]{graphics}
\usepackage[dvips]{graphicx}
\usepackage{soul}

\usepackage{color}

\numberwithin{equation}{section}
\usepackage{color}

\newcommand{\eqa}{\begin{eqnarray}}
\newcommand{\eeqa}{\end{eqnarray}}
\newcommand{\beq}{\begin{equation}}
\newcommand{\eeq}{\end{equation}}

\newtheorem{theorem}{Theorem}

\newtheorem{remark}[theorem]{Remark}

\newcommand{\benumerate}{\begin{enumerate}}
\newcommand{\eenumerate}{\end{enumerate}}

\newcommand{\bitemize}{\begin{itemize}}

\newcommand{\eitemize}{\end{itemize}}

\newcommand{\der}[2]{\frac{\partial #1}{\partial #2}}

\newcommand{\dersec}[2]{\frac{\partial^{2} #1}{\partial #2^{2}}}

\newcommand{\dermixd}[3]{\frac{\partial^{2} #1}{\partial #2 \partial #3}}

\newcommand{\ovl}[1]{\overline{#1}}

\begin{document}

\title{Shock dynamics of phase diagrams}

\author{Antonio Moro \\
\\
{\it Department of Mathematics and Information Sciences, Northumbria University} \\
{\it Newcastle upon Tyne, United Kingdom}\\
~\\
{\small antonio.moro@northumbria.ac.uk}
}
\date{}

    \maketitle


\begin{abstract}
A thermodynamic phase transition denotes a drastic change of state of a physical system due to a continuous change of thermodynamic variables, as for instance pressure and temperature. The classical van der Waals equation of state is the simplest model that predicts the occurrence of a critical point associated with the gas-liquid phase transition. 
Nevertheless, below the critical temperature theoretical predictions of the van der Waals theory significantly depart from the observed physical behaviour. We develop a novel approach to classical thermodynamics based on the solution of Maxwell relations for a generalised family of nonlocal entropy functions. This theory provides an exact mathematical description of discontinuities of the order parameter  within the phase transition region, it explains the universal form of the equations of state and the occurrence of triple points in terms of the dynamics of nonlinear shock wave fronts. \\


\end{abstract}


\section{Introduction}
A phase transition is a general concept denoting a drastic change of physical properties of a system from one state to another. Everyday experience as well as sophisticated physical experiments provide numerous examples of processes where a certain physical system undergoes a thermodynamic phase transition, e.g. boiling water, melting ice, magnetic transitions, phase separation of mixtures, superfluid He, superconductors~(see e.g. \cite{Callen,Stanley,Landau,Cyrot,Felderhof,StanleyNat}). 
More specifically, a thermodynamic phase transition is a change of state of a physical system in thermodynamic equilibrium  induced by a continuous change of thermodynamic variables as, for instance, pressure, temperature, volume. The celebrated van der Waals model provides the first simple example of equation of state that predicts the occurrence of a critical point associated with a gas-liquid phase transition.

\begin{figure}[htbp]
\begin{center}
 \includegraphics[height=6cm]{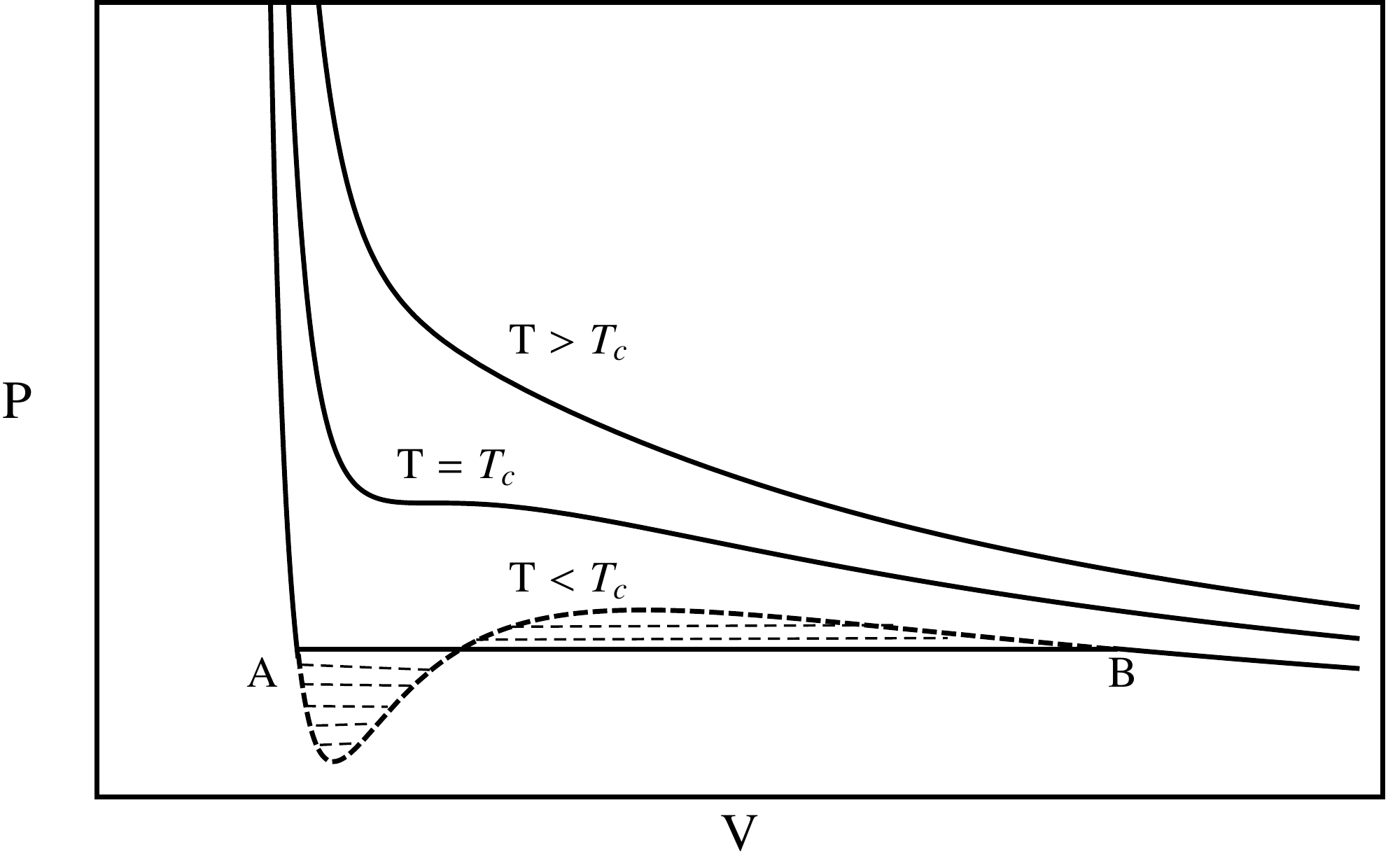}
 \end{center}
  \caption{{\footnotesize  
 Isothermal curves for a van der Waals fluid. The dashed line corresponding to the theoretical behaviour as predicted by the classical van der Waals equation is replaced by a constant pressure line according to the Maxwell equal areas rule.}}
 \label{vdwpic}
\end{figure}

Figure~\ref{vdwpic} shows the behaviour of three isothermal curves as predicted by the van der Waals equation of state. Above a certain critical temperature $T_{c}$, the curves show a smooth monotonic decrease of pressure as a function of volume. At the critical temperature the isothermal curve develops an inflection at the point $(V_{c},P_{c})$. The critical point $(V_{c},P_{c},T_{c})$ detects the occurrence of a phase transition in the system.
Below the critical temperature the van der Waals model predicts an oscillating behaviour, the dashed curve in Figure~\ref{vdwpic}, associated to a metastable state that is usually not observed. Typical experimental isothermal curves show that within the oscillating region the pressure remains constant as long as the fluid is compressed from the volume $V_{B}$ to the volume $V_{A}$.
Hence, the volume, as a function of pressure, experiences a jump that is associated with the coexistence of both gas and liquid phases. The classical criterion to resolve the discrepancy between theoretical and experimental results within the phase transition region $AB$ is known in thermodynamics as the Maxwell principle. Maxwell's criterion allows to recover the physical isotherm from the theoretical one by requesting that the intersection of the constant pressure line with the van der Waals isotherm detects two regions of equal areas as shown in Figure~\ref{vdwpic}. The theoretical justification of the Maxwell principle relies on the observation that the equal areas condition is equivalent to the fact that the Gibbs potential admits two equal valued minima across the phase transition~(see e.g. \cite{Callen}).

The approach to classical thermodynamics recently introduced in~\cite{DNM} shows
how van der Waals type equations of state can be constructed as solutions to nonlinear hyperbolic conservation laws.  Under suitable general assumptions on the functional form of the entropy, the first law of thermodynamics can be equivalently formulated in terms of a Riemann-Hopf type partial differential equation. Such equations are well known to be integrable and exactly solvable by the method of characteristics.

As a specific example, let us consider a macroscopic physical system, as for instance a fluid characterised by its Gibbs potential $\Phi$, entropy function $S$, pressure $P$, volume $V$ and temperature $T$. The first law of thermodynamics  in differential form reads as follows
\begin{equation}
\label{firstlaw}
d\Phi(P,T) = - S(P,T) dT + V(P,T) dP.
\end{equation}
For a number of models of physical interest, as for example the ideal gas, the van der Waals gas and some of its virial generalisations, the entropy function is of the form
\begin{equation}
\label{Sfun}
S = S\left(V(P,T), T\right)
\end{equation}
i.e. it is characterised by an implicit dependence on the pressure $P$ via the volume function $V(P,T)$. As shown in~\cite{DNM}, in the case of separable entropy functions of the form
\begin{equation}
\label{Sfun2}
S = \tilde{S}\left(V\right) + F(T),
\end{equation}
the first law of thermodynamics is equivalent to the Riemann-Hopf type equation
\begin{equation}
\label{hopf_type}
\der{V}{P} + \alpha(V) \der{V}{T} = 0 
\end{equation}
where $\alpha(V) :=  \tilde{S}'(V)^{-1}$. We note that~(\ref{Sfun2}) includes important classical examples such as the ideal gas entropy
\begin{equation*}
S_{\textup{id}} = N \log \left (\frac{e V}{N} \right) - N g(T)
\end{equation*}
and the van der Waals gas entropy 
\begin{equation}
\label{vdwentropy}
S_{\textup{vdw}} = N \log \left(\frac{e}{N} (V - N b) \right) - N h(T),
\end{equation}
where $g(T)$ and $h(T)$ are certain functions of the temperature (see e.g.~\cite{Landau} p. 234). For both the ideal and the van der Waals gas the equation~(\ref{hopf_type}) reduces to the Riemann-Hopf equation of the form $V_{P} + V V_{T} = 0$ (subscripts stand for the partial differentiation). We emphasise that the equation~(\ref{hopf_type}) reveals that all thermodynamic models specified by entropy functions of the form~(\ref{Sfun2}) are integrable and exactly solvable in the sense of the classical theory of hyperbolic conservation laws.

In the present paper, we introduce a generalisation of the van der Waals equation of state based on the assumption of a  weak nonlocal dependence of the entropy function on the order parameter. The main idea is to model isothermal curves within the phase transition region by smooth curves where the transition is associated with a smooth jump. 

Hence, introducing the quantity $\Delta p_{s}$, the typical size of the transition region, and $\Delta p_{d}$, the typical length of the domain under consideration, we can introduce the small non-dimensional parameter $\nu = \Delta p_{s}/\Delta p_{d}$. We also assume that isobaric curves are such that  $\nu = \Delta t_{s}/\Delta t_{d} $ where $\Delta t_{s}$ and $\Delta t_{d}$ are introduced in a similar fashion. 
\begin{figure}[htbp]
\begin{center}
\includegraphics[height=6cm]{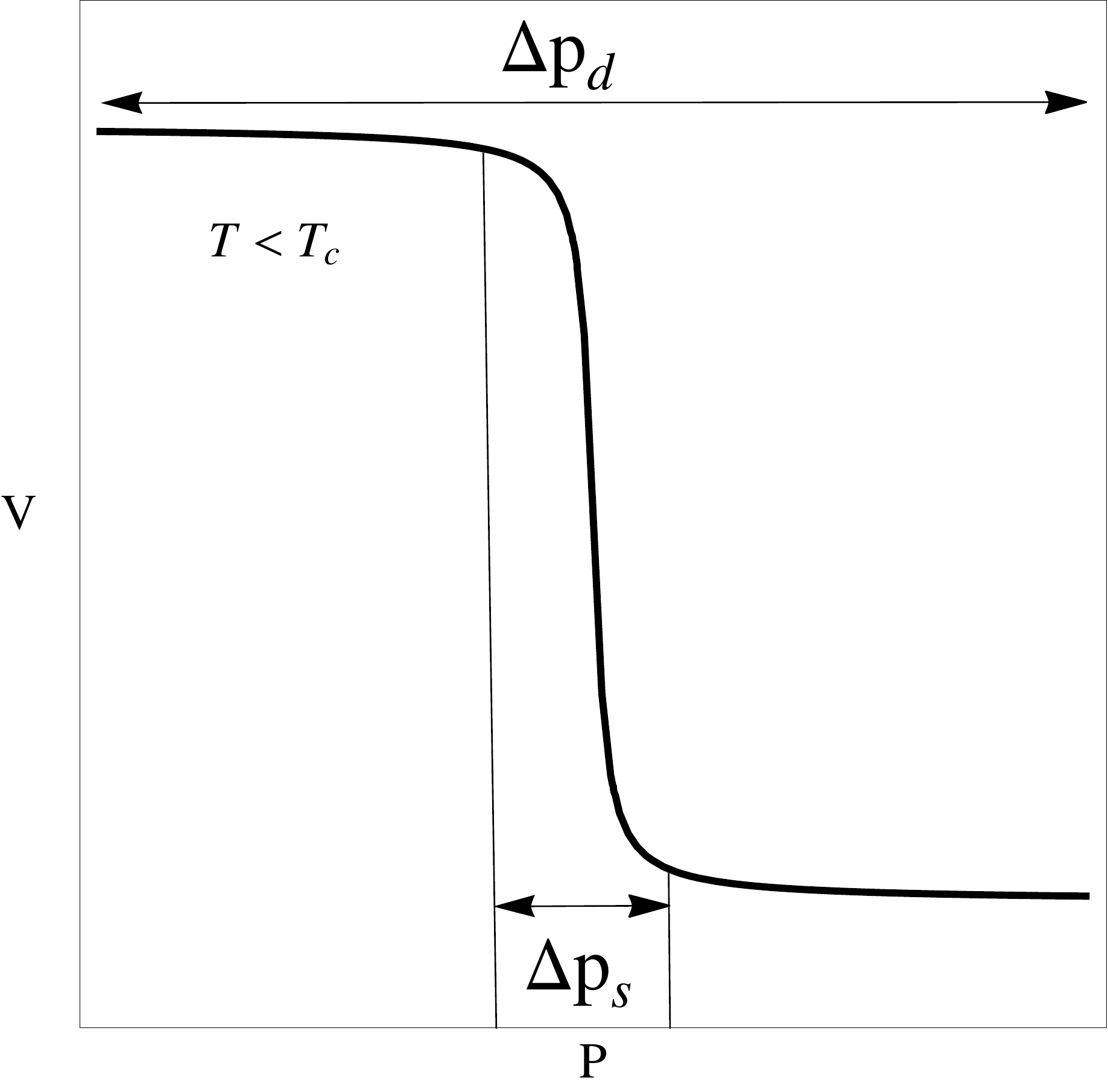} 
 \end{center}
  \caption{{\footnotesize  
The typical length of the transition region for isothermal curves is assumed to be small compared with the size of the domain under consideration, $\nu = \Delta p_{s} / \Delta p_{d} << 1$.}}
\label{pic_nu} 
\end{figure}

\noindent For, we consider the generalised entropy function defined as
\begin{equation}
\label{Sgen}
S_{\nu}(P,T) = S(V(P,T),T,\phi_{\nu}(P,T),\psi_{\nu}(P,T))
\end{equation}
where $\phi_{\nu}$ and $\psi_\nu$ are nonlocal variables defined as
\begin{equation}
\label{nonloc_var}
\phi_{\nu} := V(P+ \nu \Delta p_{d}/2 ,T) - V(P -\nu \Delta p_{d}/2 ,T) \qquad  \psi_{\nu} := V(P,T+ \nu \Delta t_{d}/2) - V(P,T-\nu \Delta t_{d}/2).
\end{equation}
The function~(\ref{Sgen}) is required to be such that in the one-phase region the limit entropy
\[
S(P,T) = \lim_{\nu \to 0} S_{\nu}(P,T)
\]
reproduces the classical results for van der Waals type systems.
The above nonlocality introduces a weak dependence of the entropy on the volume change rate. This results in a small viscous effect whose contribution becomes relevant in the two-phase region only. Consequently, far from the critical region where the volume change rate is not too large, such a dependence is negligible and the entropy reduces to the form~(\ref{Sfun}). As the system approaches the critical point and the volume change rate becomes large, nonlocal effects are no longer negligible and the dynamics of the phase diagram is governed, in a suitable double scaling regime, by the Burgers equation. In this regime, the model is integrable and exactly solvable as the Burgers equation is well known to be transformed into the heat equation via a Cole-Hopf transformation. The solution to the Burgers equation provides the required generalisation of the van der Waals equation of state and reproduces, in the zero viscosity limit, the jump (weak) solution to the Riemann-Hopf equation, the classical shock. 
Therefore, isothermal/isobaric curves are interpreted as nonlinear wave solutions to scalar conservation laws and their discontinuities associated with a phase transition find a natural description in terms of classical shocks.

This establishes an intriguing relation between thermodynamics and the theory of nonlinear conservation laws and reveals the existence of a correspondence table between phase transitions phenomenology and shock waves dynamics. 

In particular, the equal areas principle independently introduced in the theory of hyperbolic conservation laws acquires, in virtue of the above mentioned correspondence, a precise thermodynamic interpretation. We observe that the speed of a classical shock wave given by the Rankine-Hugoniot condition is identified with the Clapeyron equation which expresses the slope of the vapour-pressure curve in terms of the latent heat. Furthermore, the multiscale analysis near the critical point establishes a direct connection between the notion of universal critical behaviour of viscous breaking waves~\cite{I,DE} and the notion of universality in thermodynamics. The corresponding equation of state is given in terms of a particular solution to the Burgers equation related to the Pearcey integral. As an application we compute the critical exponents associated with the isothermal compressibility and the volume jump across the vapour-pressure curve. 

The main difference between our method and more standard approaches is that equations of state are derived via a direct integration of Maxwell's relations, for a suitable approximation of entropies of the form~(\ref{Sgen}), rather than starting from an ansatz on the asymptotic expansion of the Gibbs potential, as in the classical Landau's theory, or its scaling properties, as in Widom's theory~\cite{Stanley,Widom}.
The present approach reveals that for all models described by separable entropy functions of the form~(\ref{Sfun2}) (and their nonlocal generalisations in the family~(\ref{Sgen})) the form of the equation of state is determined by the volume depending component of the entropy, i.e. the function $\tilde{S}(V)$ in~(\ref{Sfun2}) and its nonlocal generalisations. The knowledge of a suitable finite number of isothermal/isobaric curves allows to fix uniquely the equation of state and provide a full description of isotherms and isobars in the space of intensive variables $P$ and $T$. However, the temperature depending component of the entropy, e.g. the function $F(T)$ in~(\ref{Sfun2}), determines the properties of other state curves such as isentropic curves whose convexity features have been proven to be relevant in the study of actual shock waves propagating in real fluids~\cite{Bethe,Weyl,Menikoff} (see Remarks~\ref{remark1} and~\ref{remark2}).

Finally, we observe that the Burgers dynamics allows to interpret the formation of triple points as a confluence of two classical shock waves. A discontinuity of the order parameter across the coexistence curve of two phases is interpreted as a shock wave travelling on the phase diagram. Then, according to the theory of Burgers' equation the two shocks collide in correspondence of the triple point and emerge as a single shock of increased strength. 

The analysis presented here, provided the validity of our asymptotic approximations, applies to all phenomenological and  possibly mean field theories leading to entropy functions of the form~(\ref{Sgen}). Therefore, although the present approach to thermodynamic phase transitions  is  macroscopic and phenomenological, it is natural to ask to what extent an entropy function of the form~(\ref{Sgen})  can be derived from more fundamental microscopic arguments. In fact, a vast literature is devoted to the study of statistical models (e.g. random matrix models, spin glasses) whose thermodynamic properties, in suitable critical regimes, are described by integrable nonlinear equations, see e.g.~\cite{JJ,FIK,TW,TBAZW,S,GB,BFT,CG} (this short list of references is aimed as an example and it is not meant to be complete).
For example, in~\cite{BFT, GB} it was observed that  shock wave solutions to the Burgers equation provide a description of symmetry breaking in mean field spin models.

\section{Equations of State}
\label{sec_eqstate}
The first law of thermodynamics in the form~(\ref{firstlaw}) implies the Maxwell relations $S = -\partial \Phi/ \partial T$, $V =  \partial \Phi/\partial P$. Eliminating the Gibbs potential by a further differentiation we obtain the balance equation
\begin{equation}
\label{eqbalance}
\der{V}{T} + \der{S}{P} = 0.
\end{equation}
Under the assumption~(\ref{Sfun2}) the equation~(\ref{eqbalance}) reduces to the Riemann-Hopf type equation~(\ref{hopf_type}).
The general solution $V(P,T)$ to the equation~(\ref{hopf_type}) is then obtained by the standard method of characteristics via the implicit formula
\beq
\label{hopf_sol}
T - \alpha(V) P - f(V) = 0,
\eeq
where $f(V)$ is an arbitrary function. The equation~(\ref{hopf_sol}) provides an infinite family of equations of state parametrised  by the function $f(V)$. In particular, the classical van der Waals equation for a real gas corresponds to the particular choice
\begin{equation}
\label{vdw_coeff}
\alpha(V)  =\frac{V - n b}{n R} \qquad f(V)  = \frac{n a}{V} - \frac{n^{2} b}{R V^{2}}
\end{equation}
where $n$ is the number of moles and $a$ and $b$ are constant associated, respectively, with the mean field interaction and the volume of gas particles.

 As discussed in~\cite{DNM}, a particular equation of state can be specified provided a suitable finite number of isothermal/isobaric curves is known. If, for instance, both functions $\alpha(V)$ and $f(V)$ are unknown, they can be obtained by solving the following linear system
\begin{align*}
&T_{1} - \alpha(V) P_{1}(V) - f(V) = 0 \\
&T_{2} - \alpha(V) P_{2}(V) - f(V) = 0,
\end{align*}
where the graphs of functions $P_{1}(V)$ and $P_{2}(V)$ represent any two particular isothermal curves respectively at the temperature $T_{1}$ and $T_{2}$. Functions $P_{1}(V)$ and $P_{2}(V)$ can be obtained, for example, via interpolation of experimental data. If the specific form of the entropy, and consequently the function $\alpha(V)$ is known, the above procedure applies just to a single isothermal curve and it is equivalent  to the solution of the PDE~(\ref{hopf_type}) with a particular initial datum. As long as the solution is sufficiently regular all other isotherms are uniquely determined and one can predict the occurrence of a critical point $(V_{c},P_{c},T_{c})$ defined by the following conditions
\[
T_{c} - \alpha(V_{c}) P_{c} = f(V_{c}) \qquad  - \alpha'(V_{c}) P_{c} = f'(V_{c}) \qquad - \alpha''(V_{c}) P_{c} = f''(V_{c}).
\]
These conditions mean that the critical point belongs to the critical isotherm $T_{c}$ and it is a gradient catastrophe and inflection point for the function $V=V( P,T_{c})$. It is also assumed throughout the paper that the critical point is {\it generic}, i.e. the function $f(V)$  is such that the inflection point is non-degenerate~\cite{DE}.

\begin{figure}[htbp]
\begin{center}
\includegraphics[height=6cm]{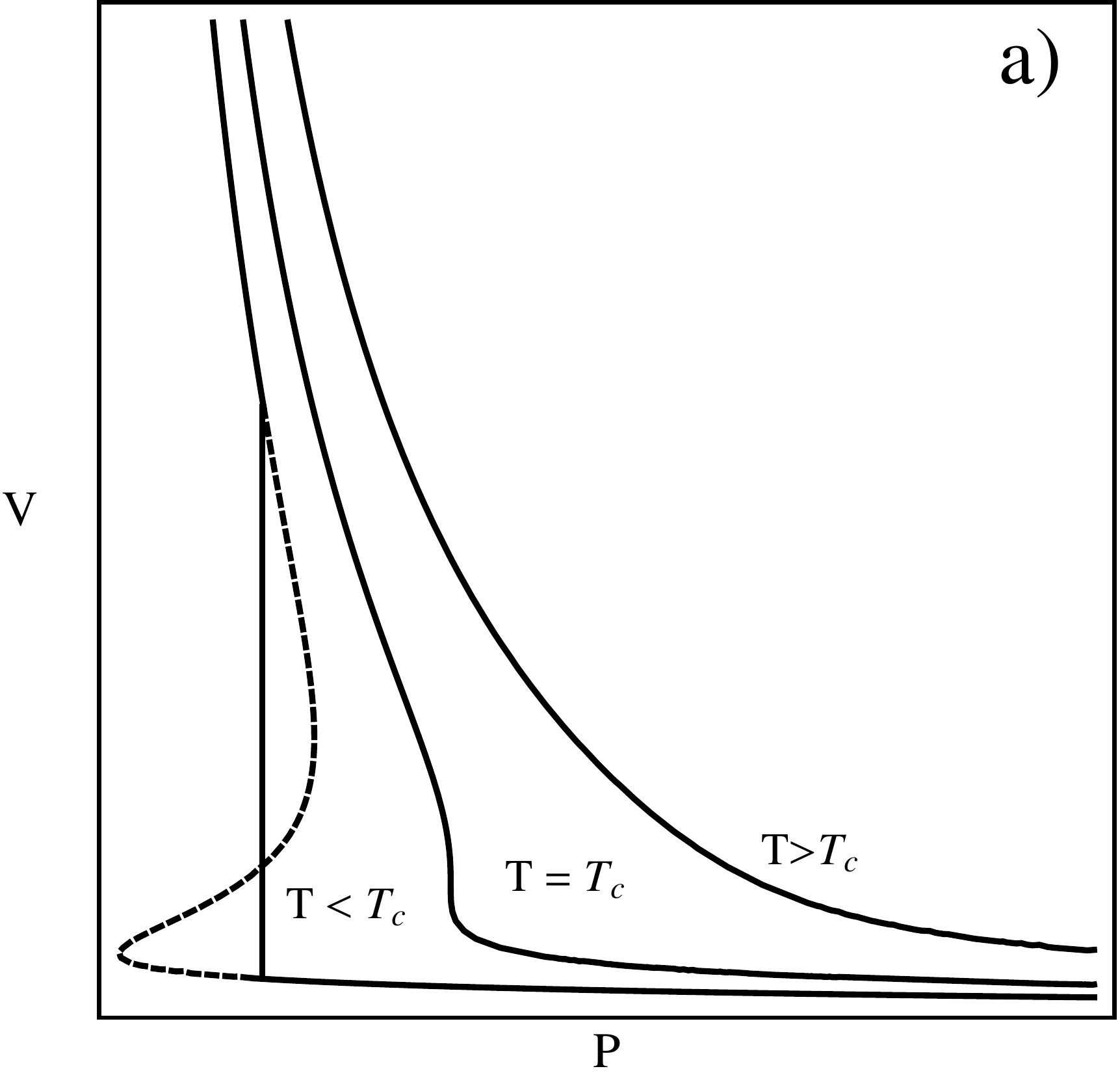} \qquad \includegraphics[height=6cm]{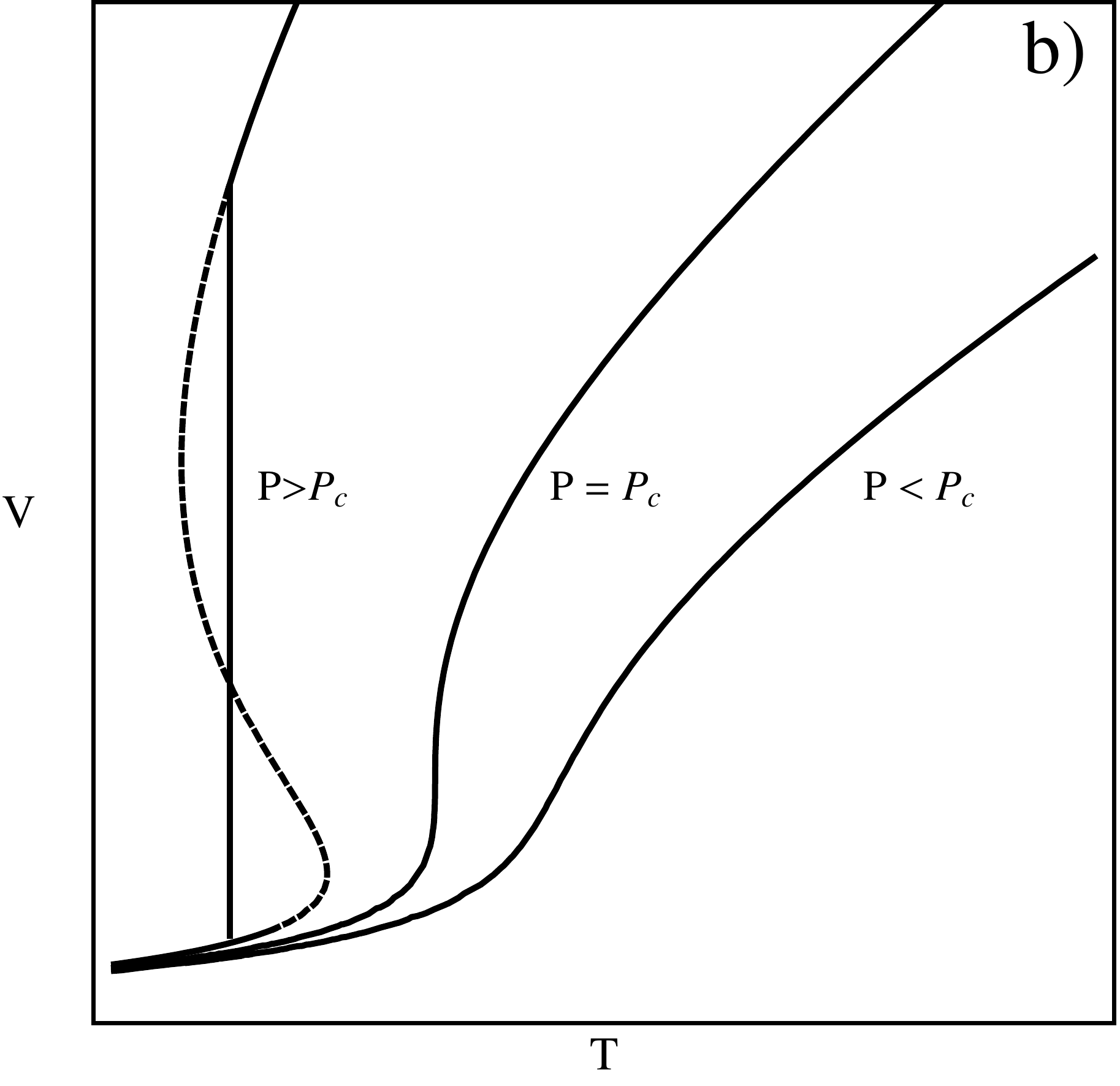}
 \end{center}
  \caption{{\footnotesize Nonlinear breaking of van der Waals a) isothermal and b) isobaric curves. The critical point corresponds to a gradient catastrophe for the function $V(P,T)$. In  the two-phase (multivalued) region experimental curves develop a jump associated with the first order phase transition.}}
\label{pic_VTVP}
\end{figure}

\begin{figure}[htbp]
\begin{center}
\includegraphics[height=8cm]{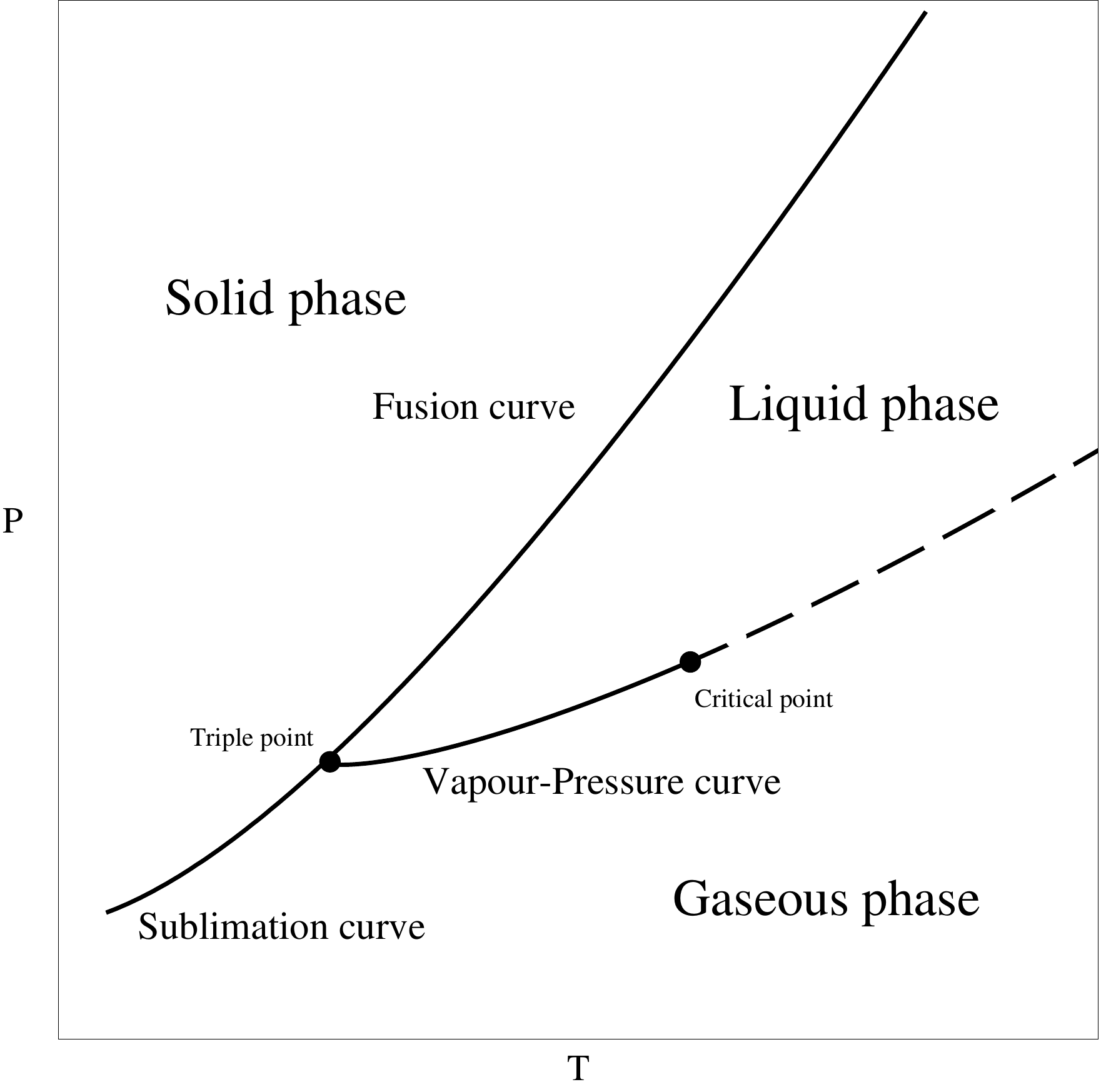}
 \end{center}
  \caption{{\footnotesize Typical phase diagram for gas-liquid-solid phase transitions.}}
  \label{triple1}
\end{figure}

Hence, isothermal curves evolve on the $PV$ space towards a phase transition just as nonlinear waves break by developing a gradient catastrophe. 
For the sake of simplicity, let us consider the van der Waals equation of state that is obtained for the particular choice~(\ref{vdw_coeff}). Figure~\ref{pic_VTVP}.$\left.\textup{a} \right)$  shows the evolution of a certain isothermal curve above the critical temperature. At the critical temperature the isotherm develops a gradient catastrophe resulting in the multivalued solution $V( P, T_{c})$. Applying a standard argument in the theory of hyperbolic nonlinear waves, since the multivalued solution is unphysical, we replace it with a discontinuous solution (the shock) according to the equal areas rule. Similarly, Figure~\ref{pic_VTVP}.$\left.\textup{b} \right)$ shows the nonlinear breaking of a typical isobaric curve.
The shock speed $U = \dot{P}(T)$, where $P = P(T)$ is the trajectory of the shock on the $PT$ plane, is given by the Rankine-Hugoniot condition
\[
U = \dot{P}(T) = \frac{\Delta S}{\Delta V}
\]
where $\Delta S$ and $\Delta V$ denote respectively the entropy and volume jumps across the coexistence curve between two phases. Recalling the definition of latent heat $L_{h} = T \Delta S$, we have
\[
\dot{P} =\frac{L_{h}}{T \Delta V}.
\]
Remarkably, we recognise that the above Rankine-Hugoniot condition is equivalent to the Clapeyron equation~\cite{Callen}. Clapeyron's equation is well known in thermodynamics as it relates the latent heat to the slope of the coexistence curve separating two phases, as for instance the vapour-pressure curve showed in Figure~\ref{triple1}.

Now, following~\cite{I,DE}, we derive the form of the equation of state in the vicinity of the critical point $(V_{c}, P_{c},T_{c})$. Introducing the displacement variables around the critical point
\begin{equation}
\label{disp_var}
\ovl{T} = \frac{T-T_{c} + \alpha(V_{c}) (P-P_{c})}{\lambda^{3}} \qquad \ovl{P} = \frac{P-P_{c}}{\lambda^{2}}  \qquad \ovl{V} = \frac{V- V_{c}}{\lambda} 
\end{equation}
where $\lambda$ is a small positive parameter and expanding the l.h.s. of the equation~(\ref{hopf_sol}) in Taylor series (for $\lambda  \to 0$) we obtain, at the leading order, the following equation
\begin{equation}
\label{expand2}
\ovl{T}-  \alpha'(V_{c}) \ovl{V} \;\ovl{P}   - \frac{1}{6} \left(f'''(V_{c}) + \alpha'''(V_{c}) P_{c} \right) \ovl{V}^{3} =0.
\end{equation}
In the original variables the local form of the solution reads
\begin{equation}
\label{scaling1}
V \simeq V_{c} + \lambda \ovl{V} (\ovl{P}, \ovl{T}) = V_{c} + \lambda \ovl{V} \left(\frac{T -T_{c} - \alpha (V_{c}) (P-P_{c})}{\lambda^{3}},  \frac{P-P_{c}}{\lambda^{2}} \right).
\end{equation}
This description of classical thermodynamics in terms of nonlinear wave equations is consistent with the local construction of the Gibbs potential via singularity theory. 
In fact, we observe that the local expansion~(\ref{expand2}) coincides with  the van der Waals equation for a real gas in a neighbourhood of the critical point and it can be interpreted as the universal unfolding of a cusp singularity ($A_{3}$ type according to Arnold's classification)~\cite{Arnold,Yung}.

\section{Phase transitions, classical shocks and universality}
We show how the proposed generalised entropy function~(\ref{Sgen}) provides a detailed description of first order phase transitions and predicts (for $\nu \to 0$) the occurrence of discontinuous isothermal/isobaric curves near the critical point. In particular, a classical result by Il'in allows to recover the universal local form of the function $V= V(P,T)$~\cite{I,DE}. We will proceed with the construction of a suitable asymptotic expansion in the small scaling parameter $\nu$. Our approach mimics to some extent the classical Landau theory in the critical region (see e.g.~\cite{Callen,Stanley}). We construct an asymptotic expansion for the entropy in terms of the gradient of the volume which plays the role of order parameter. This asymptotic expansion combined with the first law~(\ref{eqbalance}) leads to a nonlinear viscous conservation law which is approximated, after a suitable multiscale expansion, by the Burgers equation. The solution to the Burgers equation in the zero-viscosity limit allows to compute the phase diagrams in terms of a classical shock dynamics. 

Let us now consider the entropy function~(\ref{Sgen}). In the one-phase region, where the function $V(P,T)$ is assumed to be smooth, we can approximate
\[
\phi_{\nu} \simeq \nu \Delta p_{d} V_{P} \qquad \psi_{\nu} \simeq \nu \Delta t_{d} V_{T}.
\]
Therefore, we study the entropy function of the form
\begin{equation}
\label{entropy_gen}
S_{\nu}(P,T) \simeq S \left(V,T,\nu V_{P}, \nu V_{T} \right) \qquad \nu << 1.
\end{equation}
We observe that as long as the gradient of the function $V(P,T)$ is bounded the entropy function~(\ref{entropy_gen}) reduces, in the limit $\nu \to 0$, to the form~(\ref{Sfun}) and the description of the thermodynamic system in terms of the classical van der Waals theory applies. Nevertheless, the function~(\ref{entropy_gen}) is expected to produce a deviation from the classical behaviour as the system approaches the critical point where the gradient of the volume becomes large.
 
Expanding the entropy function~(\ref{entropy_gen})  at the first order in the perturbation parameter 
\begin{equation}
\label{entropyexp}
S_{\nu} \simeq S_{0}(V) + \nu S_{1}(V) V_{P} + \nu S_{2}(V) V_{T} + F(T)
\end{equation}
and substituting into the conservation laws~(\ref{eqbalance}) we arrive at the following nonlinear partial differential equation
\begin{equation}
\label{dbalance1}
V_{P} + \alpha(V) V_{T} + \nu \beta(V) V_{T}^{2} + \nu \gamma(V) V_{TT} = 0
\end{equation}
where
\[
\alpha(V) = \frac{1}{S_{0}'}, \qquad \beta(V) = \frac{S_{1}'}{(S_{0}')^{3}}  - \frac{2 S_{1} S_{0}''}{(S_{0}')^{4}}  - \frac{S_{2}'}{(S_{0}')^{2}} + \frac{S_{2} S_{0}''}{(S_{0}')^{3}}, \qquad \gamma(V) = \frac{S_{1}}{(S_{0}')} - \frac{S_{2}}{(S_{0}')^{2}}.
\]
From now on all equalities are meant to be valid up to higher orders in the expansion parameter. We observe that, within the present asymptotic approximation, all models associated with the entropy function~(\ref{entropyexp}) can be viewed as a perturbation of the Riemann-Hopf type equation~(\ref{hopf_type}). 

Introducing the local variables~(\ref{disp_var}) into the general equation~(\ref{dbalance1}) with the particular choice of the scaling parameter $\lambda = \nu^{1/4}$ and expanding all coefficients in Taylor series we obtain the Burgers equation
\begin{equation}
\label{uburgers}
u_{\ovl{P}} + \alpha_{1} u \; u_{\ovl{T}} + \gamma_{0} u_{\ovl{T} \ovl{T}} = 0,
\end{equation}
where 
\begin{align*}
&u = \ovl{V}/\sigma&  &\gamma_{0} = \gamma(V_{c})& 
&\alpha_{0} = \alpha(V_{c})&  &\alpha_{1} = \sigma \alpha'(V_{c}).&
\end{align*}
and $\sigma$ is a constant to be determined. We also assume $\alpha_{1} \gamma_{0} < 0$. This condition guarantees that the equation~(\ref{uburgers}) admits bounded solutions. Indeed, the request $\alpha_{1} \gamma_{0} < 0$ means that Burgers' equation can be mapped, via a Cole-Hopf transformation, into the heat equation with positive thermal conductivity. Hence, in terms of the original variables the solution near the critical point takes the form
\begin{equation}
\label{scaledsol}
V = V_{c} + \sigma \nu^{1/4} u \left(\frac{T-T_{c} - \alpha_{0} (P-P_{c})}{\nu^{3/4}},\frac{P-P_{c}}{\nu^{1/2}} \right)
\end{equation}
where the function $u$ solves the equation~(\ref{uburgers}). We note that in the one-phase region the zero-viscosity limit can be recovered by rescaling the variables in such a way that $\gamma_{0} \to 0$. Then, the semiclassical theory for the Burgers equation (see e.g.~\cite{Whitham}) reproduces the van der Waals theory. In the two-phase region the semiclassical  Burgers equation predicts the formation of a jump that is consistent with the equal areas rule~\cite{Whitham}.

Assuming for instance $\alpha_{1} >0$ for a finite $\gamma_{0} < 0$ the solution takes, near the critical point, the following universal form~\cite{I,DE}
\begin{equation}
\label{univ}
u(X,Y)  = -2 \der{\log \Lambda}{X}(X,Y)
\end{equation}
where $\Lambda(X,Y)$ is the Pearcey function
\[
\Lambda(X,Y) = \int_{-\infty}^{\infty}\; e^{-\frac{1}{8} \left(z^{4} - 2 Y z^{2} + 4 X z \right)} \; dz
\]
and 
\[
X = -\frac{\alpha_{1}}{\gamma_{0}}\ovl{T} \qquad Y = - \frac{\alpha_{1}^{2}}{\gamma_{0}} \ovl{P}.
\]
Note that the expression~(\ref{univ}) is exactly the Cole-Hopf transformation that reduces the Burgers equation~(\ref{uburgers}) to the heat equation $\Lambda_{Y} = \Lambda_{XX}$.

Finally, we have to require that the formula~(\ref{univ}) matches the classical solution to the Riemann-Hopf type equation in the limit $\nu \to 0$.
Applying the standard steepest descent method for the asymptotic evaluation of the Pearcey integral as $\nu \to 0$ and recalling that $u = \ovl{V}/\sigma$, the formula~(\ref{univ}) gives the cubic equation
\begin {equation}
\label{expand3}
\ovl{T} - \alpha'(V_{c}) \ovl{P} \ovl{V}  - \frac{\gamma_{0}}{\alpha'(V_{c}) \sigma^{4}} \ovl{V}^{3} = 0
\end{equation}
which is identified with the equation~(\ref{expand2}) for the following choice of the constant $\sigma$ 
\begin{equation}
\label{sig}
\sigma =\left [\frac{6 \gamma_{0}}{\alpha'(V_{c}) (f'''(V_{c}) + \alpha'''(V_{c}) P_{c})} \right]^{1/4}.
\end{equation}
Provided the constant $\sigma$ is real, the sign of the fourth root, given the critical parameter $f'''(V_{c})$,  has to be determined consistently with the condition $\alpha_{1} \gamma_{0} <0$ (see example below).
Remarkably, the equation of state~(\ref{univ}) is universal as it does not depend either on the particular functional form of the coefficients of the entropy expansion or on the global form of the equation of state. 




Hence, the notion of universality in thermodynamics is interpreted in the present context as a manifestation of the universal singular behaviour of classical shock waves near the point of gradient catastrophe. 
Let us now use the above solution to compute some relevant physical quantities as, for example, the isothermal compressibility and the volume jump across the coexistence curve  and the corresponding critical exponents associated with the scaling of thermodynamic variables~(\ref{disp_var}).
Using the formula~(\ref{univ})  the isothermal compressibility $K_{T}$ and the associated critical exponent $\gamma$ are
\[
K_{T} := - \frac{1}{V} \der{V}{P} \simeq 2 \frac{\alpha_0 \alpha_1 \sigma}{\gamma_0 V_0} \dersec{\log \Lambda}{X} \frac{1}{\nu^{1/2}} \qquad \gamma:= - \lim_{\nu \to 0} \frac{\log K_{T}}{\log \nu} = \frac{1}{2}.
\]
The volume difference between liquid and gaseous phases is approximated as
\[
V_{L} - V_{G} \simeq \nu \Delta p_{d} \der{V}{P}(P_{c},T_{c})   \simeq -2 \Delta p_{d}  \frac{\alpha_0 \alpha_1 \sigma}{\gamma_0} \left . \dersec{\log \Lambda}{X} \right |_{(X,Y) =(0,0)} \; \nu^{1/2}
\] 
and the corresponding critical exponent $\beta$ is
\[
\beta := \lim_{\nu \to 0} \frac{\log (V_{L} - V_{G})}{\log \nu} = \frac{1}{2}.
\]

{\bf Example: Hydrogen gas.}
The van der Waals model for the hydrogen gas is specified by the following parameters
\[
a =24.76 \times 10^{-3} \; m^{6} \; Pa\; mol^{-2} \qquad b =0.02661 \times 10^{-3} \; m^{3} \; mol^{-1}.
\]
Taking for instance $n = 10^{3}$ gas moles,  the critical volume is $V_{c} = 0.07983 \;m^{3}$.
Given the value of the universal gas constant $R =8.3144 \; J\; K\;mol^{-1}$, one can immediately check that 
\begin{equation}
f'''(V_{c}) = \frac{6 n (4 n b -  a R V_{c})}{V_{c}^{5}} > 0,
\end{equation}
where $f(V)$ is given in~(\ref{vdw_coeff}).
Hence, necessary condition for $\sigma$ defined by~(\ref{sig}) to be real is that  $\gamma_0 > 0$. Moreover, the request that $\alpha_{1} \gamma_0 <0$ implies that $\alpha_{1}<0$. Observing that $\alpha'(V_{c}) = 1/n >0$, this allows to fix the sign of the fourth root in the definition~(\ref{sig}), that is
\[
\sigma = - \sigma_{0}, \qquad \textup{where} \qquad \sigma_{0} = |\sigma|.
\]

{\remark{
\label{remark1}
The solution to the equation~(\ref{dbalance1}) universally given in the asymptotic regime by the formula~(\ref{scaledsol}) provides a full description of the phase diagram of the type shown in Figure~(\ref{triple1}), and in particular of isothermal and isobaric curves. We also observe that the solution to the equation~(\ref{scaledsol}) allows to study, at least in the asymptotic regime, a necessary condition for the convexity of the entropy function. Recalling that the entropy is required to be convex as a function of the intensive variables $P$ and $T$, see e.g.~\cite{Callen}, we observe that  the necessary condition for convexity
\begin{equation}
\label{Spp}
\dersec{S}{P}(P,T) \geq 0
\end{equation}
does not depend on the function $F(T)$ in~(\ref{entropyexp}) and it can be analysed in detail
provided the functions $S_{0}(V)$, $S_{1}(V)$, $S_{2}(V)$ (or in the asymptotic regime the parameters $\alpha_{0}$, $\sigma$ and $\gamma_{0}$) are known or phenomenologically determined. Nevertheless, the positivity of the Hessian
\[
\dersec{S}{P} \dersec{S}{T} - \left (\dermixd{S}{P}{T} \right)^{2} \geq 0
\]
does depend on the function $F(T)$ on which it imposes a constraint for the thermodynamic consistency of the entropy~(\ref{entropyexp}).
}

\remark{
\label{remark2}
 We also observe that both volume and temperature dependending components of entropy functions of the form~(\ref{entropyexp}) play an important role for the analysis of the propagation of shock waves in a real fluid. In particular, assuming that the solution $V =V(P,T)$ to the equation~(\ref{dbalance1}) gives a positive expansion coefficient $\partial V(P,T)/ \partial T$, the convexity of isentropic curves, in thermodynamic notation
\begin{equation}
\label{adiaconv}
\left (\dersec{P}{V} \right)_{S} >0,
\end{equation}
guarantees that the material under consideration supports the propagation of a shock wave uniquely specified via the celebrated Hugoniot equation (see~\cite{Bethe,Weyl} for more details and~\cite{Menikoff} for a review on the  subject). This classical result is known as \textup{Bethe-Weyl Theorem}.
We also note that  the condition~(\ref{adiaconv}) expressed in terms of intensive variables reads as follows
\begin{equation}
\label{adiaconv}
\left [ \der{~}{V} - \frac{\partial S /\partial V}{\partial S / \partial T} \der{~}{T} \right ]^{2} P(V,T) > 0
\end{equation}
where the function $P = P(V,T)$ is obtained by inversion of the solution $V =V(P,T)$ to the equation~(\ref{dbalance1}).
The left hand side of the condition~(\ref{adiaconv}) clearly shows that  both volume and temperature depending components of the entropy function of the form~(\ref{entropyexp}) affects the convexity of isentropic curves.
As observed in~\cite{Bethe} the condition~(\ref{adiaconv}) is generally verified for a general class of substances far from the phase transition region and it can be in principle violated when a phase transition occurs. Nevertheless, it is generally satisfied for evaporation and condensation~\cite{Bethe}.  }
}

\section{Triple point and shock confluence}
\begin{figure}[htbp]
\begin{center}
\includegraphics[height=8cm]{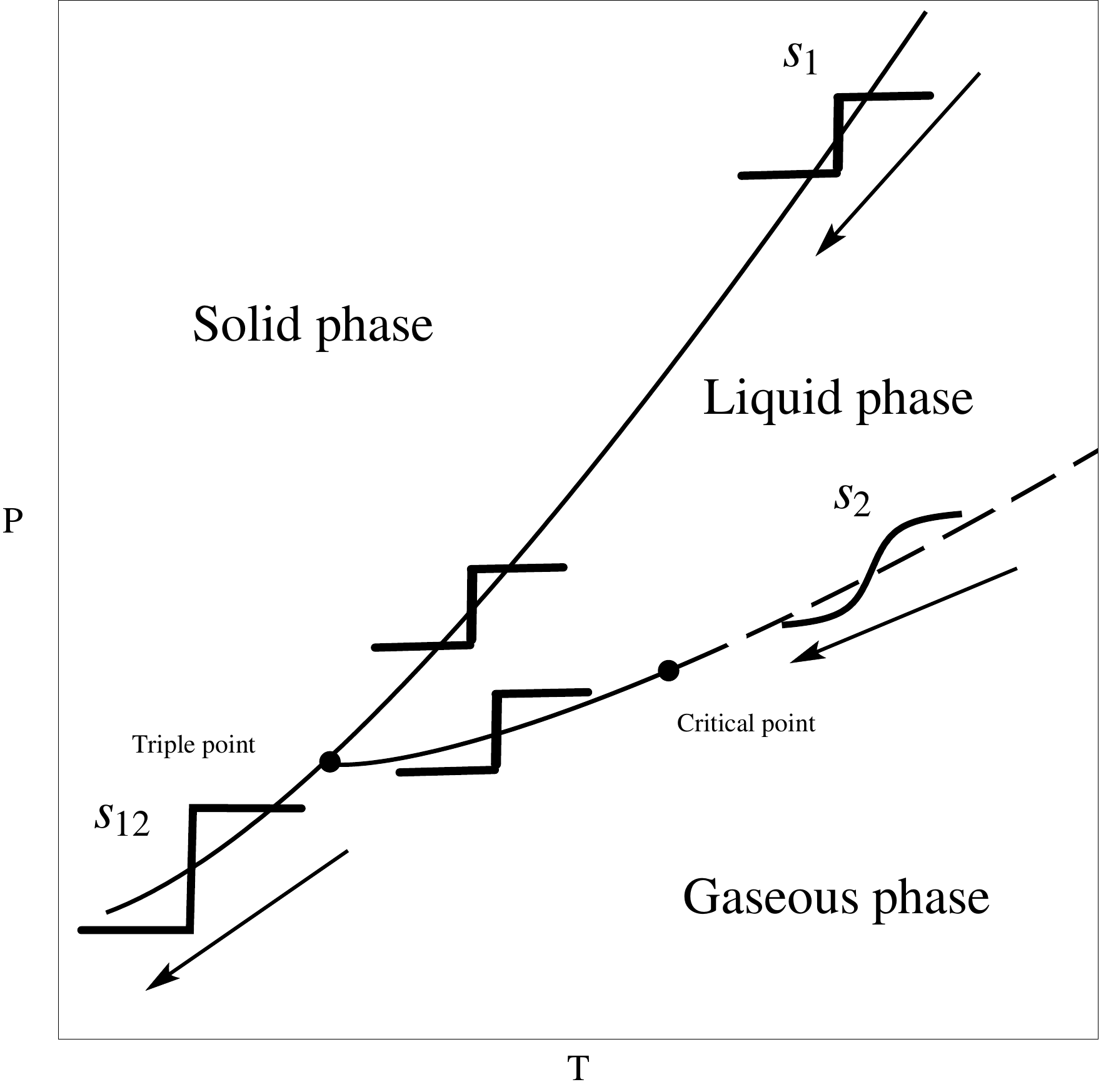}
 \end{center}
  \caption{{\footnotesize Isothermal/isobaric curves develop shock singularities across the coexistence curves between two phases, they propagate along such curves and collide at the triple point emerging as a single shock (shock confluence).}}
 \label{triple3}
\end{figure}

Let us assume that the asymptotic approximation governed by the Burgers equation~(\ref{uburgers}) is valid within a sufficiently large domain around the critical point in the $TP$ plane as shown in Figure~\ref{triple1}. Hence, we consider the dynamics of a generic isothermal/isobaric curve on the phase plane according to the zero-viscosity limit of the Burgers equation. Let us focus, for example, on an isothermal curve above the critical temperature consisting of two shocks, $s_{1}$ and $s_{2}$, whose local structure across the fusion curve and the extrapolated vapour-pressure curve is depicted in Figure~\ref{triple3}. As the temperature decreases the shock $s_{1}$  propagates along the fusion curve with a speed given by the ``Clapeyron-Rankine-Hugoniot" condition $U_{1} =(S_{l} - S_{s})/(V_{l} - V_{s})$. The ``smooth" shock $s_{2}$ propagates along the extrapolated vapour-pressure curve and develops a jump (the classical shock) when passing through the critical point. Below the critical temperature the shock $s_{2}$ propagates with the speed  $U_{2} =(S_{g} - S_{l})/(V_{g} - V_{l}) < U_{1}$. As a consequence of their speed difference the two shocks $s_{1}$ and $s_{2}$ will collide inelastically in correspondence with the triple point emerging as a single shock propagating along the sublimation curve at speed $U_{3} =(S_{g} - S_{s})/(V_{g} - V_{s})$ .

The description of the phase diagram via Burgers's equation allows to explain the occurrence of a first order phase transition and of a triple point as part of the same shock dynamics.

\section{Concluding remarks}
The semiclassical theory of nonlinear viscous equations suggests a simple, macroscopic, phenomenological generalisation of  classical van der Waals type models. This generalisation provides a detailed description of first order phase transitions  such that the structure of phase diagrams in the one-phase region and in a neighbourhood of the critical point is explained, respectively, in terms of the integrable dynamics of the Riemann-Hopf and the Burgers equation. 


Provided the system undergoes a phase transition the breaking mechanism is universal and it does not depend on the specific functional form of the coefficients in the expansion~(\ref{entropyexp}). This follows from the classical result by Il'in~\cite{I} in the case the equation~(\ref{dbalance1}) reduces or is approximated by the Burgers equation and it is expected to be valid in general according to the recent conjecture formulated by Dubrovin and Elaeva in~\cite{DE}. Moreover, as observed in~\cite{ALM}, the universal solution of the form~(\ref{scaledsol}) satisfies the second order linearisable ODE
\[
u_{XX} + 3 u u_{X} + u^{3} - Y u = X
\]
where $Y$ plays the role of a parameter. This equation can be viewed as a deformation of the var der Waals equation of state in a neighbourhood of the critical point (see e.g.~\cite{Yung}).

On the other hand all features of systems modelled by a viscous conservation law within the class considered here can be understood in terms of their thermodynamic counterpart. 

An interesting class of models which would deserve a dedicated analysis is the class of integrable viscous conservation laws that provides an exactly solvable family of models parametrised by a single function of one variable~\cite{ALM} . These models include a viscous analog of the Camassa-Holm equation that is related to the Klein-Gordon equation~\cite{ALM} and admits an interesting class of discontinuous (weak) solutions~\cite{Falqui}. 
We finally summarise our results with a first attempt to fill in a correspondence table between some fundamental concepts/objects in classical thermodynamics and their counterpart in the theory of nonlinear conservation laws, see Table~\ref{tab_correspondence}.

\begin{table}
\begin{center}
\begin{tabular}{|lcl|}

    \hline
	{\bf Thermodynamics} & & {\bf Nonlinear conservation laws}    \\ 
	\hline
Isothermal/isobaric curves  & $\leftrightarrow$ & Nonlinear waves  \\
\hline
Critical point  & $\leftrightarrow$ & Gradient catastrophe  \\
\hline
Phase transition  & $\leftrightarrow$ & Shock  \\
	\hline
Maxwell principle  & $\leftrightarrow$ & Equal areas rule  \\
\hline
Clapeyron Equation   & $\leftrightarrow$ &  Rankine-Hugoniot condition  \\
\hline
Triple point   & $\leftrightarrow$ &  Shock confluence  \\
\hline
Universality & $\leftrightarrow$ & Universality \\
\hline

\end{tabular}
\end{center}
\caption{\footnotesize Correspondence between some classical concepts/objects in thermodynamic theory and their counterpart in the theory of nonlinear conservation laws.}
\label{tab_correspondence}
\end{table}

\section*{Acknowledgements}
I wish to thank A. Arsie, A. Barra, G. De Nittis, B. Dubrovin, G. Falqui, T. Grava, Y. Kodama, P. Lorenzoni, S. Starr for stimulating discussions and useful references and the anonymous referee for his constructive comments. Thanks to G. Dossena for his useful comments on the manuscript. I am also grateful to SISSA-International School for Advanced Studies Trieste and the Department of Mathematics and Applications at University of Milano-Bicocca, where this work has been partially conceived, for their kind hospitality and the stimulating scientific environment. This research has been partially supported by the ERC grant FroM-PDE and the INDAM-GNFM Grant, Progetti Giovani 2011, prot. 50.

\end{document}